\documentclass[12pt]{article}
\usepackage[english]{babel}
\usepackage{amsmath,amsthm,amssymb,amsfonts}
\usepackage{color}
\newtheorem{theorem}{Theorem}
\newtheorem{lemma}{Lemma}
\newtheorem{definition}{Definition}
\newtheorem{remark}{Remark}
\newtheorem{example}{Example}
\newtheorem{problem}{Problem}
\begin{document}

\title{On the decoupling problem of general quasilinear first order systems\\ in two independent variables}

\author{M.~Gorgone, F.~Oliveri and M.~P.~Speciale\\
\ \\
{\footnotesize Department of Mathematical and Computer Sciences,}\\
{\footnotesize Physical Sciences and Earth Sciences, University of Messina}\\
{\footnotesize Viale F. Stagno d'Alcontres 31, 98166 Messina, Italy}\\
{\footnotesize mgorgone@unime.it; foliveri@unime.it; mpspeciale@unime.it}
}

\date{Published in \textit{J. Math. Anal. Appl.} \textbf{446}, 276--298 (2017).}

\maketitle
   
\begin{abstract}
The paper deals with the decoupling problem of general quasilinear first order systems in two independent variables. We 
consider either the case of homogeneous and autonomous systems or the one of nonhomogeneous and/or nonautonomous 
systems. Necessary and sufficient conditions for the partial or full decoupling of the systems at hand are provided. The 
conditions involve the properties of eigenvalues and eigenvectors of the coefficient matrix, and provide the differential 
constraints whose integration leads to the decoupling transformation. Some applications of physical interest are also given.
\end{abstract}



\section{Introduction}
Many physical problems are mathematically modeled by first order systems of partial differential equations in the form of balance laws \cite{Bressan,Dafermos}, that in one space dimension read as
\begin{equation}
\label{balance}
\frac{\partial\mathbf{f}^0(\mathbf{u})}{\partial t}+\frac{\partial\mathbf{f}^1(\mathbf{u})}{\partial x}=\mathbf{g}(\mathbf{u}),
\end{equation}
where $\mathbf{u}\in \mathbb{R}^n$ denotes the unknown vector field, 
$\mathbf{f}^0(\mathbf{u})\in\mathbb{R}^n$ collects the components of the densities of some physical quantities, $\mathbf{f}^1(\mathbf{u})\in\mathbb{R}^n$ the components of the corresponding fluxes, and $\mathbf{g}(\mathbf{u})\in\mathbb{R}^n$ the production 
terms; when $\mathbf{g}(\mathbf{u})\equiv\mathbf{0}$, we have a system of conservation laws. 

Systems like (\ref{balance}) fall in the more general class of nonhomogeneous
quasilinear first order systems of partial differential equations,
\begin{equation}
\label{quasilinear_auto}
A^0(\mathbf{u})\frac{\partial \mathbf{u}}{\partial t}+ A^1(\mathbf{u})\frac{\partial\mathbf{u}}{\partial x}=\mathbf{g}(\mathbf{u}),
\end{equation}
where $A^0(\mathbf{u})$ and  $A^1(\mathbf{u})$ are $n\times n$ matrices (the gradient matrices of $\mathbf{f}^0(\mathbf{u})$ and $\mathbf{f}^1(\mathbf{u})$, respectively, in the case of conservative systems).

The analytical, as well as numerical, treatment of quasilinear systems of conservation laws is 
in general a difficult task. In the case of hyperbolic systems, 
the generalized eigenvalues of the matrix pair $\{A^0,A^1\}$, giving the wave
speeds, depend on $\mathbf{u}$, whereupon the shape of the various components in the
solution will vary in time: rarefaction waves will decay, and compression waves will
become steeper, possibly leading to shock formation in finite time \cite{Bressan,Dafermos}. 
Since also the eigenvectors, determining the approximate change of field variables across a wave, depend on $\mathbf{u}$, 
nontrivial interactions between different waves will occur; the strength of the interacting waves may change, and new waves of
different families can be created, as a result of the interaction.

For a  strictly hyperbolic system of conservation laws we have $n$ families
of waves, each corresponding to an eigenvalue of the system. The nonlinearity of wavespeeds
leads to the formation of shocks, so that solutions must be understood in the weak
sense. The existence and stability of global weak solutions for Cauchy data with small
total variation was established by Glimm  \cite{Glimm}. For systems of more than two equations nonlocal resonant
interaction effects  between different families of waves are observed, leading to a variety of new phenomena \cite{Majda}
such as  blowup of solutions \cite{Hunter}, and delay in the onset of shocks \cite{Majda2}. The resonance determines the occurrence of solutions exhibiting a strong nonlinear instability in the form of catastrophic blowup of solutions \cite{Hunter,Joly}. In fact, there are systems where Cauchy data with arbitrarily small oscillation can grow arbitrarily large in finite time \cite{Young1993,Young2003,Young2006}.

Special problems of physical interest (see \cite{Cristescu,Jeffrey1982,RogersRuggeri1985,Anliker1971,Pedley1980,CurroOliveri2008,Oliveri2012} for some examples)
may require to consider systems where the coefficients may depend also on the 
independent variables $(t,x)$, accounting for material inhomogeneities, or 
particular geometric assumptions, or external actions, so that in some applications one may need to consider 
nonautonomous and/or nonhomogeneous quasilinear systems of the form
\begin{equation}
\label{quasilinear_nonauto}
A^0(t,x,\mathbf{u})\frac{\partial \mathbf{u}}{\partial t}+ A^1(t,x,\mathbf{u})\frac{\partial\mathbf{u}}{\partial x}=\mathbf{g}(t,x,\mathbf{u}).
\end{equation}

In dealing with quasilinear systems, it may be interesting to look for the conditions (if any) leading to their possible decoupling into smaller non--interacting subsystems (full decoupling), or their reduction to a set of smaller subsystems that can be solved separately in hierarchy (partial decoupling).  

For homogeneous and autonomous first order quasilinear systems of partial differential equations in two independent variables, the decoupling problem can be formulated as follows \cite{Bogo2,Bogo3}.
\begin{problem}
When can a  system like 
\begin{equation}\label{Courant_system}
\frac{\partial {u}_\ell}{\partial t}=\sum_{j=1}^n A_{\ell j}(u_1,\ldots,u_n)\frac{\partial u_j}{\partial x}, \quad \ell=1,\ldots,n,
\end{equation}
be locally decoupled in some coordinates $v_1(\mathbf{u}),\ldots , v_n(\mathbf{u})$ into $k$ 
non-interacting subsystems, say
\begin{equation}
\frac{\partial v_{m_j+i}}{\partial t}=\sum_{\ell=1}^{n_j}\widetilde{A}_{m_j+i,m_j+\ell}(v_{m_j+1},\ldots,v_{m_j+n_j})\frac{\partial v_{m_j+\ell}}{\partial x},
\end{equation}
of some orders $n_1$, \ldots, $n_k$ with $n_1+\ldots+n_k=n$, where $j=1,\ldots,k$, $i=1,\ldots,n_j$,  and $m_j=n_1+\ldots+n_j$?
\end{problem}

A first result has been obtained by Nijenhuis \cite{Nijenhuis} in the case of a strictly hyperbolic system; the necessary and sufficient conditions for the complete decoupling of system (\ref{Courant_system}) into $n$ non--interacting one--dimensional subsystems require the vanishing of the corresponding Nijenhuis tensor
\begin{equation}
N_{jik}=A_{\alpha i}\frac{\partial A_{jk}}{\partial u_\alpha}-A_{\alpha k}\frac{\partial A_{ji}}{\partial u_\alpha}
+A_{j\alpha}\frac{\partial A_{\alpha  i}}{\partial u_k}-A_{j \alpha}\frac{\partial A_{\alpha k}}{\partial u_i}.
\end{equation}

The decoupling problem has been considered by Bogoyavlenskij \cite{Bogo2,Bogo3}, who provided necessary and sufficient conditions by using  Nijenhuis \cite{Nijenhuis} and Haantjes \cite{Haantjes} tensors. More in detail, to reduce system (\ref{Courant_system}) into block-diagonal form with $k$ mutually interacting blocks  of dimensions $n_i\times n_j$ \cite{Bogo2,Bogo3}
it is necessary and sufficient that in the tangent spaces $T_\mathbf{x}(\mathbb{R}^n)$ there exist $k$ smooth distributions
$L_{1\mathbf{x}} , \ldots , L_{k\mathbf{x}}$ of dimensions $n_1, \ldots , n_k$ such that
$L_{1\mathbf{x}}\oplus\cdots\oplus L_{k\mathbf{x}}=T_\mathbf{x}(\mathbb{R}^n)$ and the conditions
\begin{equation}
A(L_{i\mathbf{x}})\subset L_{i\mathbf{x}}, \qquad N(L_{i\mathbf{x}},L_{i\mathbf{x}})\subset L_{i\mathbf{x}}, \qquad
N(L_{i\mathbf{x}},L_{j\mathbf{x}})\subset L_{i\mathbf{x}}+L_{j\mathbf{x}},
\end{equation}
hold provided that the eigenvalues of the operator $A$ in any two different subspaces $L_{i\mathbf{x}}$ and $L_{j\mathbf{x}}$  are different almost everywhere for $\mathbf{x}\in\mathbb{R}^n$ ($i\neq j$; $i, j \in \{1, \ldots , k\}$).
In the more restrictive case of the decoupling into $k$ non--interacting blocks, the different blocks
$ \widetilde{A}_{m_j+i,m_j+\ell}$ depend on different variables; hence, for the generic case the eigenvalues corresponding to any two blocks
$ \widetilde{A}_{m_j+i,m_j+\ell}$ do not coincide with each other almost everywhere for $\mathbf{x}\in\mathbb{R}^n$ (while inside a given block some eigenvalues can coincide). In this case, the necessary and sufficient conditions for the reducibility of the systems (\ref{Courant_system})  into $k$
non-interacting subsystems have the form 
\begin{equation}
A(L_{i\mathbf{x}})\subset L_{i\mathbf{x}}, \qquad N(L_{i\mathbf{x}},L_{i\mathbf{x}})\subset L_{i\mathbf{x}}, \qquad
N(L_{i\mathbf{x}},L_{j\mathbf{x}})=0.
\end{equation}

Within this theoretical framework, a couple of recent papers by Tunitsky \cite{Tunitsky1,Tunitsky2}, who established necessary and sufficient conditions for transforming quasilinear first order systems into block triangular systems by using a geometric formalism for such equations, based on Nijenhuis and Haantjes tensors, are worth of being quoted.  

In this paper, we shall consider either autonomous and homogeneous first order quasilinear systems like 
\begin{equation}\label{system1}
\frac{\partial \mathbf{u}}{\partial t}+A(\mathbf{u})\frac{\partial \mathbf{u}}{\partial x}=\mathbf{0},
\end{equation}
or general nonhomogeneous and/or nonautonomous ones, say
\begin{equation}\label{system2}
\frac{\partial \mathbf{u}}{\partial t}+A(t,x,\mathbf{u})\frac{\partial \mathbf{u}}{\partial x}=\mathbf{g}(t,x,\mathbf{u})
\end{equation}
(possibly coming from systems in conservative form),
and obtain the necessary and sufficient conditions allowing for the partial decoupling in two or more subsystems, as we shall precise below.   When such a \emph{partial} decoupling is possible,  we may solve the various subsystems separately in hierarchy. Also, we shall prove how to extend the conditions to be satisfied in order to characterize the systems that can be fully decoupled into non--interacting subsystems. 
Differently from Bogoyavlenskij's approach,  the conditions we shall discuss later involve, as one expects, just the properties of the eigenvalues, the eigenvectors (together with the generalized eigenvectors, if needed) of the coefficient matrix; in particular, the conditions for the full decoupling of a
hyperbolic system in $k$ non--interacting subsystems require the vanishing both of the change of characteristic speeds of a subsystem across a wave of the other subsystems, and of the interaction coefficients between waves of different subsystems. Even if the computation of eigenvalues and eigenvectors of the coefficient matrix may be hard (especially for large matrices), the
conditions we derived have a simple interpretation, as we shall discuss later.  
Moreover,  when the required decoupling conditions are satisfied, we have also the 
differential constraints whose integration provides the variable transformation leading to the (partially or fully) decoupled system. The results presented in this paper arose from a generalization of the ones found by a direct approach in the case of the decoupling problem for quasilinear first order systems involving two or three dependent variables \cite{GOS_AAPP}.

The plan of the paper is the following. In Section~\ref{main1}, after introducing the 
definitions of partially and fully decoupled systems, we establish the necessary and sufficient 
conditions for the decoupling of hyperbolic first order homogeneous and autonomous quasilinear systems. Then, 
in Section~\ref{main2}, we extend this result to general first order homogeneous and autonomous quasilinear systems. 
In Section~\ref{main3},
we face the decoupling problem of general nonhomogeneous and/or nonautonomous 
systems. Finally, in Section~\ref{applications}, some examples of systems of physical 
interest that can be, under suitable conditions, partially or fully decoupled are presented.

\section{Decoupling of hyperbolic homogeneous and autonomous quasilinear systems}\label{main1}
In this Section, we consider the case of a hyperbolic first order homogeneous and autonomous quasilinear system of partial differential equations in two independent variables, and provide the necessary and sufficient conditions for decoupling it.

Let us introduce the notation that will be used throughout the paper, and define the meaning of partially and fully decoupled systems. 

\begin{definition}[Notation]
Given $\mathbf{U}\equiv (U_1,U_2,\ldots,U_n)^T\in\mathbb{R}^n$, and a set of $k\ge 2$ integers $n_1,\ldots,n_k$ such that
$n_1+\ldots+n_k=n$,  let us relabel and group the components of $\mathbf{U}$ as follows:
\begin{equation}
\label{groupU}
\left\{\{U^{(1,1)},\ldots,U^{(1,n_1)}\},\ldots,\{U^{(k,1)},\ldots,U^{(k,n_k)}\}\right\}.
\end{equation}
Moreover, let us set
\begin{equation}
\begin{aligned}
\mathcal{U}_i=\bigcup_{r=1}^i \left\{U^{(r,1)},\ldots, U^{(r,n_r)}\right\},\qquad
\overline{\mathcal{U}}_i=\bigcup_{r=i+1}^k\left\{U^{(r,1)},\ldots,U^{(r,n_r)}\right\};
\end{aligned}
\end{equation}
the cardinality of the set $\mathcal{U}_i$ is $m_i$, 
whereas the cardinality of the set $\overline{\mathcal{U}}_i$ is $n-m_i$, where $m_i=n_1+\ldots+n_i$.
\end{definition}

\begin{definition}[Partially decoupled systems]
\label{def:partialk}
The first order quasilinear system 
\begin{equation}\label{systemU}
\frac{\partial \mathbf{U}}{\partial t}+T(\mathbf{U})\frac{\partial \mathbf{U}}{\partial x}=\mathbf{0},
\end{equation}
$T$  being an $n\times n$ real matrix with entries smooth functions depending on $\mathbf{U}$, 
is partially decoupled in $2\le k\le n$ subsystems of some orders $n_1,\ldots, n_k$ $(n_1+\ldots+n_k = n)$
if, relabelling and suitably collecting the components of $\mathbf{U}$ in $k$ subgroups as in (\ref{groupU}),
we recognize $k$ subsystems such that the $i$-th subsystem $(i=1,\ldots, k)$
involves at most the $m_i$ field variables of the set $\mathcal{U}_i$.
\end{definition}

\begin{definition}[Fully decoupled systems]
\label{def:fullk}
The first order quasilinear system (\ref{systemU})
is fully decoupled in $2\le k\le n$ subsystems of some orders $n_1,\ldots, n_k$ $(n_1+\ldots+n_k=n)$
if, relabelling and suitably collecting the components of $\mathbf{U}$ in $k$ subgroups as in (\ref{groupU}), 
we recognize $k$ subsystems such that the $i$-th subsystem $(i=1,\ldots, k)$
involves exactly the $n_i$ field variables $\{U^{(i,1)},\ldots,U^{(i,n_i)}\}$.
\end{definition}

The following lemma  will lead us to prove a theorem providing necessary and sufficient conditions 
for the partial decoupling of a hyperbolic first order quasilinear system in two independent variables.

\begin{lemma}\label{lemma:partialk}
Let $T$ be an $n\times n$ lower triangular block real matrix, say
\begin{equation}
\label{matrix_general}
T=
\left[
\begin{array}{llllll}
T^1_1 & 0^1_2 &0^1_3 & \dots & 0^1_{k-1} & 0^1_k\\
T^2_1 & T^2_2 & 0^2_3 & \dots & 0^2_{k-1} & 0^2_k\\
\dots & \dots & \dots & \dots &\dots &\dots\\
\dots & \dots & \dots & \dots &\dots &\dots\\
T^{k-1}_1 & T^{k-1}_2 & T^{k-1}_3 & \dots & T^{k-1}_{k-1} & 0^{k-1}_k\\
T^k_1 & T^k_2 & T^k_3 & \dots & T^k_{k-1} & T^k_k
\end{array}
\right],
\end{equation}
$T^i_j$ being $n_i\times n_j$ matrices with entries smooth functions depending on $\mathbf{U}\equiv (U_1,\ldots,U_n)$, and 
$0^i_j$  $n_i\times n_j$ matrices of zeros $(n_1+\ldots +n_k=n,\, 2\le k\le n)$.
Let us assume that matrix $T$ has real eigenvalues and a complete set of eigenvectors. 
The entries of matrices $T^i_j$ $(i=1,\ldots,k,\,j=1,\ldots,i)$ 
depend at most on the $m_i$ variables of the set $\mathcal{U}_i$
if and only if:
\begin{enumerate}
\item the set of the eigenvalues of $T$ (counted with their multiplicity) and the  corresponding left and right eigenvectors can be divided
into $k$ subsets each containing $n_i$ $(i=1,\ldots,k)$ elements, 
\begin{equation}
\label{lambdaLR}
\begin{aligned}
&\left\{\{\Lambda^{(1,1)},\ldots,\Lambda^{(1,n_1)}\},\ldots,\{\Lambda^{(k,1)},\ldots,\Lambda^{(k,n_k)}\}\right\},\\
&\left\{\{\mathbf{L}^{(1,1)},\ldots,\mathbf{L}^{(1,n_1)}\},\ldots,\{\mathbf{L}^{(k,1)},\ldots,\mathbf{L}^{(k,n_k)}\}\right\},\\
&\left\{\{\mathbf{R}^{(1,1)},\ldots,\mathbf{R}^{(1,n_1)}\},\ldots,\{\mathbf{R}^{(k,1)},\ldots,\mathbf{R}^{(k,n_k)}\}\right\},
\end{aligned}
\end{equation}
where 
\[
\{\Lambda^{(i,1)},\ldots,\Lambda^{(i,n_i)}\}
\]
are the eigenvalues (counted with their multiplicity) of matrix $T^i_i$; 
\item the following structure conditions hold true:
\begin{equation}\label{struct_general}
\begin{aligned}
&\left(\nabla_{\mathbf{U}}\Lambda^{(i,\alpha)}\right)\cdot \mathbf{R}^{(j,\gamma)}=0,\\
&\mathbf{L}^{(i,\alpha)}\cdot\left((\nabla_{\mathbf{U}}\mathbf{R}^{(\ell,\beta)}) \mathbf{R}^{(j,\gamma)}\right)=0, \\
&i=1,\ldots, k-1, \; \ell=1,\ldots,i,\\ 
&\alpha=1,\ldots,n_i,\;\beta=1,\ldots,n_\ell, \;\alpha\neq\beta\;\hbox{if}\;i=\ell,\\
&j=i+1,\ldots,k, \; \gamma=1,\ldots,n_j, 
\end{aligned}
\end{equation}
where 
\[
\nabla_{\mathbf{U}}\equiv\left(\frac{\partial}{\partial U^{(1,1)}},\ldots,\frac{\partial}{\partial U^{(1,n_1)}},\ldots,\frac{\partial}{\partial U^{(k,1)}},\ldots,\frac{\partial}{\partial U^{(k,n_k)}}\right).
\]
\end{enumerate}
\end{lemma}
\begin{proof}
Let the lower triangular block matrix $T$ be such that  the entries of matrices $T^i_j$ 
($i=1,\ldots,k-1$, $j=1,\ldots,i$) 
depend at most on the elements of the set $\mathcal{U}_i$.

The set of the $n$ eigenvalues of $T$ is the union of the sets of the $n_i$ eigenvalues of $T^i_i$ $(i=1,\ldots, k)$;   since the entries of matrix $T^i_i$ depend at most on the elements of the set $\mathcal{U}_i$, the same can be said for its $n_i$ eigenvalues.

Let us denote the set of left and right eigenvectors of matrix $T$  as in (\ref{lambdaLR}), and
group the components of a right (left, respectively) eigenvector $\mathbf{R}^{(r,\alpha)}$ 
($\mathbf{L}^{(r,\alpha)}$, respectively) as follows:
\[
\mathbf{R}^{(r,\alpha)}=\left(\begin{array}{c}
\mathbf{R}^{(r,\alpha)}_1\\
\mathbf{R}^{(r,\alpha)}_2\\
\ldots\\
\mathbf{R}^{(r,\alpha)}_k
\end{array}
\right),\qquad
\mathbf{L}^{(r,\alpha)}=\left(
\mathbf{L}^{(r,\alpha)}_1,
\mathbf{L}^{(r,\alpha)}_2,
\ldots,\\
\mathbf{L}^{(r,\alpha)}_k\right),
\] 
where $\mathbf{R}^{(r,\alpha)}_i$ ($\mathbf{L}^{(r,\alpha)}_i$, respectively) are column (row, respectively) vectors with $n_i$ components. 

Taking into account the relations for the left eigenvectors,
\begin{equation}
\label{rel_left}
\begin{aligned}
\mathbf{L}_1^{(r,\alpha)}T^1_1+\mathbf{L}_2^{(r,\alpha)}T^2_1+\ldots+
\mathbf{L}_{k-1}^{(r,\alpha)}T^{k-1}_1+&\mathbf{L}_k^{(r,\alpha)}T^k_1=\Lambda^{(r,\alpha)}\mathbf{L}_1^{(r,\alpha)},\\
\mathbf{L}_2^{(r,\alpha)}T^2_2+\ldots+
\mathbf{L}_{k-1}^{(r,\alpha)}T^{k-1}_2+&\mathbf{L}_k^{(r,\alpha)}T^k_2=\Lambda^{(r,\alpha)}\mathbf{L}_2^{(r,\alpha)},\\
&\ldots\ldots\\
\mathbf{L}_{k-1}^{(r,\alpha)}T^{k-1}_{k-1}+&\mathbf{L}_k^{(r,\alpha)}T^k_{k-1}=\Lambda^{(r,\alpha)}\mathbf{L}_{k-1}^{(r,\alpha)},\\
&\mathbf{L}_k^{(r,\alpha)}T^k_k=\Lambda^{(r,\alpha)}\mathbf{L}_k^{(r,\alpha)},
\end{aligned}
\end{equation}
since $\Lambda^{(r,\alpha)}$ is an eigenvalue of $T^r_r$, if $r<k$, we can choose $\mathbf{L}^{(r,\alpha)}_{r+1}, \ldots, \mathbf{L}^{(r,\alpha)}_{k}$
as zero row vectors; this means that 
the left eigenvectors $\mathbf{L}^{(r,\alpha)}$ $(\alpha=1,\ldots,n_r)$ may have non--vanishing only the first $m_r$ components $(m_{r}=n_1+\ldots +n_r)$;  moreover, due to the hypotheses of the functional dependence of matrices $T^i_j$, the components of $\mathbf{L}^{(r,\alpha)}_{1},\ldots,\mathbf{L}^{(r,\alpha)}_r$
depend at most on the elements of the set $\mathcal{U}_r$.

Analogously, by considering the relations for the right eigenvectors,
\begin{equation}
\label{rel_right}
\begin{aligned}
&T^1_1\mathbf{R}^{(r,\alpha)}_1=\Lambda^{(r,\alpha)}\mathbf{R}^{(r,\alpha)}_1,\\
&T^2_1\mathbf{R}^{(r,\alpha)}_1+T^2_2\mathbf{R}^{(r,\alpha)}_2=\Lambda^{(r,\alpha)}\mathbf{R}^{(r,\alpha)}_2,\\
&\ldots\ldots\\
&T^{k-1}_1\mathbf{R}^{(r,\alpha)}_1+T^{k-1}_2\mathbf{R}^{(r,\alpha)}_2+\ldots+
T^{k-1}_{k-1}\mathbf{R}^{(r,\alpha)}_{k-1}=\Lambda^{(r,\alpha)}\mathbf{R}^{(r,\alpha)}_{k-1},\\
&T^k_1\mathbf{R}^{(r,\alpha)}_1+T^k_2\mathbf{R}^{(r,\alpha)}_2+\ldots+
T^k_{k-1}\mathbf{R}^{(r,\alpha)}_{k-1}+T^k_k\mathbf{R}^{(r,\alpha)}_k=\Lambda^{(r,\alpha)}\mathbf{R}^{(r,\alpha)}_k,
\end{aligned}
\end{equation}
since $\Lambda^{(r,\alpha)}$ is an eigenvalue of $T^r_r$, if $r>1$, we can choose $\mathbf{R}^{(r,\alpha)}_{1}, \ldots, \mathbf{R}^{(r,\alpha)}_{r-1}$
as zero column vectors; this means  that the right eigenvectors $\mathbf{R}^{(r,\alpha)}$ $(\alpha=1,\ldots,n_r)$
for $r>1$ may have non--vanishing only the last $n-m_{r-1}$ components; moreover, due to the hypotheses of the functional dependence of matrices $T^r_r$, the components of $\mathbf{R}^{(r,\alpha)}_s$ $(s=r,\ldots,k)$ depend at most on the elements of the set $\mathcal{U}_s$. Notice that, because of the hyperbolicity assumption, the vectors
$\mathbf{R}^{(r,\alpha)}_r$, as well as the vectors $\mathbf{L}^{(r,\alpha)}_r$ $(r=1,\ldots,k,\; \alpha=1,\ldots,n_r)$ are linearly independent.

As a consequence,  conditions (\ref{struct_general}) are trivially satisfied.
In fact, at most the first $m_i$ components of the vector $\nabla_\mathbf{U}\Lambda^{(i,\alpha)}$ may be non--vanishing, whereas the
first $m_{j-1}$ components of $\mathbf{R}^{(j,\gamma)}$ are zero: since $j>i$,
\begin{equation}
\left(\nabla_{\mathbf{U}}\Lambda^{(i,\alpha)}\right)\cdot \mathbf{R}^{(j,\gamma)}=0.
\end{equation}
Moreover, since the first $m_{\ell-1}$ components of $\mathbf{R}^{(\ell,\beta)}$
are vanishing, the components of $\mathbf{R}^{(\ell,\beta)}_s$ $(s=\ell,\ldots,k)$ depend at most on $\mathcal{U}_s$, the first $m_{j-1}$  of $\mathbf{R}^{(j,\gamma)}$
are vanishing, and $j>\ell$, it follows that the first $m_j$ components of the
vector $(\nabla_{\mathbf{U}}\mathbf{R}^{(\ell,\beta)}) \mathbf{R}^{(j,\gamma)}$ are vanishing; therefore, it is
\begin{equation}
\mathbf{L}^{(i,\alpha)}\cdot\left((\nabla_{\mathbf{U}}\mathbf{R}^{(\ell,\beta)}) \mathbf{R}^{(j,\gamma)}\right)=0.
\end{equation}

Viceversa, if conditions (\ref{struct_general}) hold true, then it can be proved that all entries of matrices $T^i_j$  $(i=1,\ldots,k-1,\,j=1,\ldots,i)$ depend at most on the elements of the set $\mathcal{U}_i$. 

At first, let us prove that from (\ref{struct_general})$_1$ it follows that 
$\Lambda^{(r,\alpha)}$ $(1\le r< k,\; \alpha=1,\ldots,n_r)$ can at most depend on the elements of the set $\mathcal{U}_r$.  

Let us denote with $\Lambda$ one of the eigenvalues of the matrix $T^r_r$ $(1\le r<k)$, and let us set
\begin{equation}
\nabla_{\mathbf{U}}\equiv\left(\nabla_{1},\ldots,\nabla_{k}\right),
\end{equation}
where 
\[
\nabla_{i}\equiv\left(\frac{\partial}{\partial U^{(i,1)}},\ldots,\frac{\partial}{\partial U^{(i,n_i)}}\right),\qquad i=1,\ldots,k.
\]
Since $\mathbf{R}^{(j,\gamma)}$ for $j>r$ may have non--vanishing only the last $n-m_{j-1}$ components, conditions (\ref{struct_general})$_1$ read
\begin{equation*}
\begin{aligned}
\left(\nabla_{r+1}
\Lambda\right)\cdot \mathbf{R}_{r+1}^{(r+1,\gamma)}+\left(\nabla_{r+2}\Lambda\right)\cdot \mathbf{R}_{r+2}^{(r+1,\gamma)}+\ldots+\left(\nabla_{k}\Lambda\right)\cdot \mathbf{R}_k^{(r+1,\gamma)}&=0,\\
\left(\nabla_{r+2}\Lambda\right)\cdot \mathbf{R}_{r+2}^{(r+2,\gamma)}+\ldots+\left(\nabla_{k}\Lambda\right)\cdot \mathbf{R}_k^{(r+2,\gamma)}&=0,\\ 
&\ldots\\
\left(\nabla_{k}\Lambda\right)\cdot \mathbf{R}_k^{(k,\gamma)}&=0,
\end{aligned}
\end{equation*}
whereupon it immediately follows that
\begin{equation}
\frac{\partial\Lambda}{\partial U^{(r+1,1)}}=\ldots=\frac{\partial\Lambda}{\partial U^{(r+1,n_{r+1})}}=\ldots=\frac{\partial\Lambda}{\partial U^{(k,1)}}=\ldots=\frac{\partial\Lambda}{\partial U^{(k,n_k)}}=0.
\end{equation}

Moreover, because of the lower triangular block structure of matrix $T$, the left eigenvectors $\mathbf{L}^{(i,\alpha)}$ may have non--vanishing only the first $m_i$ components, and the right eigenvectors $\mathbf{R}^{(j,\gamma)}$ may have non--vanishing only the last $n-m_{j-1}$ components, (\ref{struct_general})$_2$ can be written as:
\begin{equation}
\label{reduced}
\begin{aligned}
\sum_{r=\ell}^i\left(\mathbf{L}_r^{(i,\alpha)}\cdot\left((\nabla_{j}\mathbf{R}_r^{(\ell,\beta)}) \mathbf{R}_j^{(j,\gamma)}+
\ldots+(\nabla_{k}\mathbf{R}_r^{(\ell,\beta)}) \mathbf{R}_k^{(j,\gamma)}\right)\right)=0,
\end{aligned}
\end{equation}
for $i=1,\ldots,k-1$, $\ell \le i$, $j=i+1,\ldots,k$, and $\alpha\neq \beta$ for $\ell=i$.

From the relations
\begin{equation}
T^r_r\mathbf{R}^{(r,\alpha)}_r=\Lambda^{(r,\alpha)}\mathbf{R}^{(r,\alpha)}_r,\qquad r=1,\ldots,k-1,
\end{equation}
for $j>r$, we obtain
\begin{equation}
\left(\sum_{s=j}^k(\nabla_s T^r_r)\mathbf{R}_s^{(j,\gamma)}\right)\mathbf{R}^{(r,\alpha)}_r+(T^r_r-\Lambda^{(r,\alpha)}\mathbb{I}_r)\left(\sum_{s=j}^k(\nabla_s \mathbf{R}_r^{(r,\alpha)})\mathbf{R}_s^{(j,\gamma)}\right)=\mathbf{0}, 
\end{equation}
$\mathbb{I}_r$ being the identity $r\times r$ matrix, whereupon
\begin{equation}
\begin{aligned}
&\mathbf{L}^{(r,\beta)}_r\cdot\left(\sum_{s=j}^k(\nabla_s A^r_r)\mathbf{R}_s^{(j,\gamma)}\right)\mathbf{R}^{(r,\alpha)}_r\\
&\qquad+(\Lambda^{(r,\beta)}-\Lambda^{(r,\alpha)})\mathbf{L}^{(r,\beta)}_r\cdot\left(\sum_{s=j}^k(\nabla_s \mathbf{R}_r^{(r,\alpha)})\mathbf{R}_s^{(j,\gamma)}\right)=0.
\end{aligned}
\end{equation}
As a consequence, either when $\alpha=\beta$, or using (\ref{reduced}) when $\alpha\neq \beta$, it is
\begin{equation}
\label{a^r_r}
\mathbf{L}^{(r,\beta)}_r\left(\sum_{s=j}^k(\nabla_s T^r_r)\mathbf{R}_s^{(j,\gamma)}\right)\mathbf{R}^{(r,\alpha)}_r =0
\end{equation}
for $\alpha,\beta=1,\ldots,n_r$. Therefore, using (\ref{a^r_r}) for $j=k, k-1,\ldots, r+1$, it remains proved that all entries of the matrix $T^r_r$ $(r=1,\ldots,k-1)$, and so all the components of $\mathbf{R}_r^{(r,\alpha)}$, depend at most on the elements of the set $\mathcal{U}_r$.

Let us now take the relations (\ref{reduced}); by specializing them neatly for $\ell=i-1,i-2,\ldots,1$ and $j=k,k-1,\ldots,i+1$, it is immediately deduced that the components of $\mathbf{R}_s^{(r,\alpha)}$ for $s\ge r$ depend at most on the elements of the set $\mathcal{U}_s$.

Finally,  from  the relations
\begin{equation}
\sum_{s=r}^i T^i_s\mathbf{R}^{(r,\alpha)}_s=\Lambda^{(r,\alpha)}\mathbf{R}^{(r,\alpha)}_i, \qquad r=1,\ldots,i,\quad i<k,
\end{equation}
for $j>i$, we obtain
\begin{equation}
\label{reducedbis}
\sum_{s=r}^i \left( \left(\sum_{t=j}^k(\nabla_t T^i_s)\mathbf{R}^{(j,\gamma)}_t\right)\mathbf{R}^{(r,\alpha)}_s\right)=\mathbf{0}.
\end{equation}

By neatly specializing the relations (\ref{reducedbis}) for $r=i-1,i-2,\ldots,1$ and $j=k,k-1,\ldots,i+1$,
it follows that the entries of the matrices $T^i_s$ $(s=1,\ldots,i)$ depend at most on the $m_i$ variables of the set $\mathcal{U}_i$, and this completes the proof.

\end{proof}

\begin{remark}
Relations (\ref{struct_general}) provide  $\displaystyle\sum_{i=1}^k n_i m_i(n-m_i)$ constraints, and this is exactly the number of conditions required to ensure that the entries of matrices $T^i_j$ $(i=1,\ldots,k,\; j=1,\ldots,i)$
are independent of the elements of the set $\overline{\mathcal{U}}_i$.  
In fact, the number of entries of the matrices $T^i_j$ are $n_im_i$, and the cardinality of the set
$\overline{\mathcal{U}}_i$ is $n-m_i$. 
\end{remark} 

\begin{remark}
If matrix $T$ has the lower triangular block structure (\ref{matrix_general}) then, since the first $m_{j-1}$ components of $\mathbf{R}^{(j,\gamma)}$ are vanishing, and $j>i\geq\ell$, the first $m_i$ components of the
vector $(\nabla_{\mathbf{U}}\mathbf{R}^{(j,\gamma)}) \mathbf{R}^{(\ell,\beta)}$ 
can not be different from zero; therefore, it is identically
\[
\begin{aligned}
&\mathbf{L}^{(i,\alpha)}\cdot\left(
(\nabla_{\mathbf{U}}\mathbf{R}^{(j,\gamma)}) \mathbf{R}^{(\ell,\beta)}\right)=0,\\
&i=1,\ldots, k-1, \; \ell=1,\ldots,i,\\ 
&\alpha=1,\ldots,n_i,\;\beta=1,\ldots,n_\ell, \;\alpha\neq\beta\;\hbox{if}\;i=\ell,\\
&j=i+1,\ldots,k, \; \gamma=1,\ldots,n_j.
\end{aligned}
\]

Consequently, conditions (\ref{struct_general}) may be written as well as
\begin{equation}
\begin{aligned}
&\left(\nabla_{\mathbf{U}}\Lambda^{(i,\alpha)}\right)\cdot \mathbf{R}^{(j,\gamma)}=0,\\
&\mathbf{L}^{(i,\alpha)}\cdot\left((\nabla_{\mathbf{U}}\mathbf{R}^{(\ell,\beta)}) \mathbf{R}^{(j,\gamma)}-
(\nabla_{\mathbf{U}}\mathbf{R}^{(j,\gamma)}) \mathbf{R}^{(\ell,\beta)}\right)=0, \\
&i=1,\ldots, k-1, \; \ell=1,\ldots,i,\\ 
&\alpha=1,\ldots,n_i,\;\beta=1,\ldots,n_\ell, \;\alpha\neq\beta\;\hbox{if}\;i=\ell,\\
&j=i+1,\ldots,k, \; \gamma=1,\ldots,n_j.
\end{aligned}
\end{equation}
This result reveals useful in the proof of next theorem.
\end{remark}

By using Lemma~\ref{lemma:partialk}, it is immediately proved the following theorem.

\begin{theorem}\label{thm:partialk}
The first order quasilinear system (\ref{system1}),
assumed to be hyperbolic in the $t$--direction,
can be transformed by a smooth (locally) invertible transformation 
\begin{equation}
\label{transfk}
\mathbf{u}=\mathbf{h}(\mathbf{U}),\quad \hbox{or, equivalently,}\quad
\mathbf{U}=\mathbf{H}(\mathbf{u}),
\end{equation}
into a system like
\begin{equation}\label{targetk}
\frac{\partial\mathbf{U}}{\partial t}+T(\mathbf{U})\frac{\partial\mathbf{U}}{\partial x}=\mathbf{0},
\end{equation}
in the unknowns (\ref{groupU}), where 
$T=(\nabla_{\mathbf{u}}\mathbf{H})\,A\,(\nabla_{\mathbf{u}}\mathbf{H})^{-1}$ is  a lower triangular block matrix having the form (\ref{matrix_general})
with $T^i_j$ $(i=1,\ldots,k,\,j=1,\ldots,i)$  $n_i\times n_j$ matrices such that their entries are smooth functions depending at most on the elements of the set $\mathcal{U}_i$, 
whereas $0_j^i$ are $n_i\times n_j$ matrices of zeros, respectively, if and only if:
\begin{enumerate}
\item the set of the eigenvalues of matrix $A$ (counted with their multiplicity), and the associated left and right eigenvectors can be divided
into $k$ subsets each containing $n_i$ $(i=1,\ldots,k)$ elements
\begin{equation}\label{lambdalr}
\begin{aligned}
&\left\{\{\lambda^{(1,1)},\ldots,\lambda^{(1,n_1)}\},\ldots,\{\lambda^{(k,1)},\ldots,\lambda^{(k,n_k)}\}\right\},\\
&\left\{\{\mathbf{l}^{(1,1)},\ldots,\mathbf{l}^{(1,n_1)}\},\ldots,\{\mathbf{l}^{(k,1)},\ldots,\mathbf{l}^{(k,n_k)}\}\right\},\\
&\left\{\{\mathbf{r}^{(1,1)},\ldots,\mathbf{r}^{(1,n_1)}\},\ldots,\{\mathbf{r}^{(k,1)},\ldots,\mathbf{r}^{(k,n_k)}\}\right\};
\end{aligned}
\end{equation}
\item  the following  structure conditions hold true:
\begin{equation}\label{structk}
\begin{aligned}
&\left(\nabla_{\mathbf{u}}\lambda^{(i,\alpha)}\right)\cdot \mathbf{r}^{(j,\gamma)}=0, \\
&\mathbf{l}^{(i,\alpha)}\cdot\left((\nabla_{\mathbf{u}}\mathbf{r}^{(\ell,\beta)}) \mathbf{r}^{(j,\gamma)}-
(\nabla_{\mathbf{u}}\mathbf{r}^{(j,\gamma)}) \mathbf{r}^{(\ell,\beta)}\right)=0, \\
&i=1,\ldots, k-1, \; \ell=1,\ldots,i,\\ 
&\alpha=1,\ldots,n_i,\;\beta=1,\ldots,n_\ell, \;\alpha\neq\beta\;\hbox{if}\;i=\ell,\\
&j=i+1,\ldots,k, \; \gamma=1,\ldots,n_j.
\end{aligned}
\end{equation}
\end{enumerate}
Moreover, the decoupling variables $U^{(i,\alpha)}=H^{(i,\alpha)}(\mathbf{u})$ 
$(i=1,\ldots,k-1, \,\alpha=1,\ldots,n_i)$ are found from 
\begin{equation}
\left(\nabla_\mathbf{u}H^{(i,\alpha)}\right)\cdot \mathbf{r}^{(j,\gamma)}=0,
\end{equation}
where $j=i+1,\ldots,k, \, \gamma=1,\ldots,n_j$.
\end{theorem}

\begin{proof}
Let us consider the  hyperbolic system (\ref{system1}),
and denote the  $k$ subsets each containing $n_i$ $(i=1,\ldots,k)$ eigenvalues (counted with their multiplicity), together with their associated left and right eigenvectors as in (\ref{lambdalr}). The hyperbolicity condition implies that the eigenvalues $\lambda^{(i,\alpha)}$ $(i=1,\ldots,k,\;\alpha=1,\ldots, n_i)$ are real, whereas the corresponding left (right, respectively) eigenvectors are linearly independent and span  $\mathbb{R}^n$.

Let us assume that  the conditions 
(\ref{structk}) are satisfied. Then, by introducing a smooth (locally) invertible transformation like (\ref{transfk}) such that
\begin{equation}
\begin{aligned}
&\left(\nabla_\mathbf{u}H^{(i,\alpha)}\right)\cdot \mathbf{r}^{(j,\gamma)}=0,\\
&i=1,\ldots,k-1,\,\alpha=1,\ldots,n_i,\\
&j=i+1,\ldots,k,\,\gamma=1,\ldots,n_j,
\end{aligned}
\end{equation}
we obtain the system (\ref{targetk}),  where $T$ is a lower triangular block matrix like 
(\ref{matrix_general}).

It remains to prove that the entries of the matrices $T^i_j$ $(i=1,\ldots,k-1,\,j=1,\ldots,i)$ do not depend on the elements of the set $\overline{\mathcal{U}}_i$.

It is
\begin{equation}\label{prop1}
\lambda^{(i,\alpha)}=\Lambda^{(i,\alpha)}, \qquad
\mathbf{l}^{(i,\alpha)}=\mathbf{L}^{(i,\alpha)}(\nabla_\mathbf{u}\mathbf{H}), \qquad
\mathbf{r}^{(i,\alpha)}=(\nabla_\mathbf{u}\mathbf{H})^{-1}\mathbf{R}^{(i,\alpha)},
\end{equation}
and also
\begin{equation}\label{prop2}
\nabla_\mathbf{u}(\cdot)=\nabla_\mathbf{U}(\cdot)(\nabla_\mathbf{u}\mathbf{H}).
\end{equation}

As a consequence, we have:
\begin{equation}
\begin{aligned}
0&=\left(\nabla_\mathbf{u}\lambda^{(i,\alpha)}\right)\cdot \mathbf{r}^{(j,\gamma)}=\\
&=\left(\nabla_\mathbf{U}\Lambda^{(i,\alpha)}\right) (\nabla_\mathbf{u}\mathbf{H})(\nabla_\mathbf{U}\mathbf{H})^{-1}  \mathbf{R}^{(j,\gamma)}=\\
&=\left(\nabla_\mathbf{U}\Lambda^{(i,\alpha)}\right)\cdot \mathbf{R}^{(j,\gamma)},  
\end{aligned}
\end{equation}
whereupon 
\begin{equation}
\label{cond1k}
\left(\nabla_\mathbf{u}\lambda^{(i,\alpha)}\right)\cdot \mathbf{r}^{(j,\gamma)}=0\quad\Leftrightarrow\quad
\left(\nabla_\mathbf{U}\Lambda^{(i,\alpha)}\right)\cdot \mathbf{R}^{(j,\gamma)}=0.
\end{equation}

Furthermore, it is
\begin{equation}\label{condLRk}
\begin{aligned}
0&=\mathbf{l}^{(i,\alpha)}\cdot\left((\nabla_{\mathbf{u}}\mathbf{r}^{(\ell,\beta)}) \mathbf{r}^{(j,\gamma)}-
(\nabla_{\mathbf{u}}\mathbf{r}^{(j,\gamma)}) \mathbf{r}^{(\ell,\beta)}\right)=\\
&=\mathbf{L}^{(i,\alpha)}\left(\nabla_\mathbf{u}\mathbf{H}\right)\left(
\nabla_\mathbf{u}\left((\nabla_{\mathbf{u}}\mathbf{H})^{-1}\mathbf{R}^{(\ell,\beta)}\right)(\nabla_{\mathbf{u}}\mathbf{H})^{-1}\mathbf{R}^{(j,\gamma)}\right.\\
&\left.\qquad\qquad-\nabla_\mathbf{u}\left((\nabla_{\mathbf{u}}\mathbf{H})^{-1}\mathbf{R}^{(j,\gamma)}\right)(\nabla_{\mathbf{u}}\mathbf{H})^{-1}\mathbf{R}^{(\ell,\beta)}\right)=\\
&=\mathbf{L}^{(i,\alpha)}\left(\nabla_\mathbf{u}\mathbf{H}\right)\left(
\nabla_\mathbf{U}\left((\nabla_{\mathbf{u}}\mathbf{H})^{-1}\mathbf{R}^{(\ell,\beta)}\right)(\nabla_{\mathbf{u}}\mathbf{H})(\nabla_{\mathbf{u}}\mathbf{H})^{-1}\mathbf{R}^{(j,\gamma)}\right.\\
&\left.\qquad\qquad-\nabla_\mathbf{U}\left((\nabla_{\mathbf{u}}\mathbf{H})^{-1}\mathbf{R}^{(j,\gamma)}\right)(\nabla_{\mathbf{u}}\mathbf{H})(\nabla_{\mathbf{u}}\mathbf{H})^{-1}\mathbf{R}^{(\ell,\beta)}\right)=\\
&=\mathbf{L}^{(i,\alpha)}\left(\nabla_\mathbf{u}\mathbf{H}\right)\left(
\nabla_\mathbf{U}\left((\nabla_{\mathbf{u}}\mathbf{H})^{-1}\mathbf{R}^{(\ell,\beta)}\right)\mathbf{R}^{(j,\gamma)}\right.\\
&\left.\qquad\qquad-\nabla_\mathbf{U}\left((\nabla_{\mathbf{u}}\mathbf{H})^{-1}\mathbf{R}^{(j,\gamma)}\right)\mathbf{R}^{(\ell,\beta)}\right)=\\
&=\mathbf{L}^{(i,\alpha)}(\nabla_\mathbf{u}\mathbf{H})\left(
\nabla_\mathbf{U}\left((\nabla_{\mathbf{u}}\mathbf{H})^{-1}\right)\left(\mathbf{R}^{(\ell,\beta)}\mathbf{R}^{(j,\gamma)}-\mathbf{R}^{(j,\gamma)}\mathbf{R}^{(\ell,\beta)}\right)\right.\\
&\left.\qquad\qquad+(\nabla_{\mathbf{u}}\mathbf{H})^{-1}\left((\nabla_{\mathbf{U}}\mathbf{R}^{(\ell,\beta)})\mathbf{R}^{(j,\gamma)}-(\nabla_{\mathbf{U}}\mathbf{R}^{(j,\gamma)})\mathbf{R}^{(\ell,\beta)}\right)\right)=\\
&=\mathbf{L}^{(i,\alpha)}\cdot\left((\nabla_{\mathbf{U}}\mathbf{R}^{(\ell,\beta)}) \mathbf{R}^{(j,\gamma)}-
(\nabla_{\mathbf{U}}\mathbf{R}^{(j,\gamma)}) \mathbf{R}^{(\ell,\beta)}\right)=\\
&=\mathbf{L}^{(i,\alpha)}\cdot\left((\nabla_{\mathbf{U}}\mathbf{R}^{(\ell,\beta)}) \mathbf{R}^{(j,\gamma)}\right),
\end{aligned}
\end{equation}
whereupon
\begin{equation}
\label{cond2k}
\begin{aligned}
&\mathbf{L}^{(i,\alpha)}\cdot\left((\nabla_{\mathbf{U}}\mathbf{R}^{(\ell,\beta)}) \mathbf{R}^{(j,\gamma)}\right)=0 \quad \Leftrightarrow\\
&\quad \Leftrightarrow\quad \mathbf{l}^{(i,\alpha)}\cdot\left((\nabla_{\mathbf{u}}\mathbf{r}^{(\ell,\beta)}) \mathbf{r}^{(j,\gamma)}-(\nabla_{\mathbf{u}}\mathbf{r}^{(j,\gamma)}) \mathbf{r}^{(\ell,\beta)}\right)=0.
\end{aligned}
\end{equation}
As a result, by using Lemma \ref{lemma:partialk}, the entries of the matrices  $T^i_j$ $(i=1,\ldots,k-1,\,j=1,\ldots,i)$ do not depend on  the elements of the set $\overline{\mathcal{U}}_i$. 

Viceversa, if the entries of the matrices $T^i_j$ $(i=1,\ldots,k-1,\,j=1,\ldots,i)$ do not depend on the elements of the set $\overline{\mathcal{U}}_i$,
\emph{i.e.}, the system is partially decoupled in $k$ subsystems,
then, because of (\ref{cond1k}) and (\ref{cond2k}), conditions (\ref{structk}) must hold, and this  concludes the proof.
\end{proof}

As a byproduct of Theorem~\ref{thm:partialk} we may recover immediately the conditions for the full decoupling problem in the case of hyperbolic first order quasilinear systems.

\begin{theorem}[Full decoupling of hyperbolic systems]\label{thm:full}
For a hyperbolic system of first order homogeneous and autonomous quasilinear 
PDEs like (\ref{system1})
to be locally reducible into $k$ non--interacting subsystems of some orders $n_1,\ldots, n_k$, with $n_1+\ldots+n_k=n$, 
in the unknowns (\ref{groupU}),
it is necessary and sufficient that:
\begin{enumerate}
\item the characteristic velocities (counted with their multiplicity), and the
corresponding left and right eigenvectors can be divided
into $k$ subsets each containing $n_i$ $(i=1,\ldots,k)$ elements as in (\ref{lambdalr});
\item  the following structure conditions hold true:
\begin{equation}
\label{mycond}
\begin{aligned}
&\left(\nabla_{\mathbf{u}}\lambda^{(i,\alpha)}\right)\cdot \mathbf{r}^{(j,\gamma)}=0, \\
&\mathbf{l}^{(i,\alpha)}\cdot\left((\nabla_{\mathbf{u}}\mathbf{r}^{(i,\beta)}) \mathbf{r}^{(j,\gamma)}-
(\nabla_{\mathbf{u}}\mathbf{r}^{(j,\gamma)}) \mathbf{r}^{(i,\beta)}\right)=0, \\
&\forall\,i,j=1,\ldots, k,\quad i\neq j,\\
&\alpha,\beta=1,\ldots,n_i, \quad \alpha\neq\beta, \quad \gamma=1,\ldots,n_j.
\end{aligned}
\end{equation}
\end{enumerate}
Moreover, the decoupling variables 
\[
U^{(i,\alpha)}=H^{(i,\alpha)}(\mathbf{u}), 
\]
are found from 
\begin{equation}
\left(\nabla_\mathbf{u}H^{(i,\alpha)}\right)\cdot \mathbf{r}^{(j,\gamma)}=0,
\end{equation}
where
\[
i,j=1,\ldots,k, \quad i\neq j,\quad \alpha=1,\ldots,n_i,\quad \gamma=1,\ldots,n_j.
\]
The coefficient matrix for a fully decoupled system results in block diagonal form (diagonal if $k=n$).
\end{theorem}
\begin{proof}
It immediately follows from Theorem~\ref{thm:partialk}.
\end{proof}

Some comments about the decoupling conditions of previous theorem and some well known facts concerned to
wave solutions of hyperbolic quasilinear systems are in order. For such systems, it is relevant to quantify the dependence of the wave speeds (the eigenvalues of 
coefficient matrix) upon the field variables (see \cite{Young1993}). More precisely, we may compute
the change of the characteristic speed $\lambda^{(i)}$
across a wave with speed $\lambda^{(j)}$, say  $\nabla_\mathbf{u}\lambda^{(i)}\cdot \mathbf{r}^{(j)}$. For $j=i$ we have
the decay coefficient $\nabla_\mathbf{u}\lambda^{(i)}\cdot \mathbf{r}^{(i)}$; this is of special importance since it determines the genuine nonlinearity or linear degeneracy of the wave \cite{Lax,Boillat}. 
In the case of completely exceptional systems, \emph{i.e.}, systems where all admitted waves are linearly degenerate,
weak waves do not give rise to shock formation.
Another important effect, which can have dramatic consequences on the behavior of solutions, is due to the interaction of two waves leading to the formation of waves in other families. In particular, for incident waves belonging to different families, 
the reflected wave is determined to leading order by the interaction
coefficient $\mathbf{l}^{(i)}\cdot\left((\nabla_{\mathbf{u}}\mathbf{r}^{(j)}) \mathbf{r}^{(\ell)}-
(\nabla_{\mathbf{u}}\mathbf{r}^{(\ell)}) \mathbf{r}^{(i)}\right)$ \cite{Young2002}.

Therefore, the conditions (\ref{mycond}), guaranteeing the decoupling of a hyperbolic first order quasilinear  system in $k$
non--interacting subsystems, have the following (obvious) meaning:
\begin{enumerate}
\item the change in the characteristic speeds of a subsystem across a wave of a different subsystem must be vanishing;
\item waves of different subsystems do not interact.
\end{enumerate}

\section{Decoupling of general homogeneous and autonomous quasilinear systems}\label{main2}
In Section~\ref{main1}, we considered the decoupling problem for hyperbolic homogeneous and autonomous quasilinear systems. Here, we investigate the case where the coefficient matrix does not possess a complete set of eigenvectors and/or has complex-valued eigenvalues. Also in this general case we give necessary and sufficient conditions for the partial or full decoupling.

\begin{definition}
Let $A$ be an $n\times n$ real matrix whose entries are smooth functions depending on $\mathbf{u}\in\mathbb{R}^n$. 
If the matrix $A$ has not a complete set of eigenvectors and/or has complex-valued eigenvalues, let us associate: 
\begin{itemize}
\item to each real eigenvalue its (left and right) 
eigenvectors and, if needed, its generalized (left and right) eigenvectors in such a way we have as many linearly independent vectors as the multiplicity of the eigenvalue;
\item to each couple of conjugate complex eigenvalues the real part and the imaginary part of its (left and right) eigenvectors (or generalized eigenvectors, if needed)
in such a way we have as many couples of linearly independent vectors as the multiplicity of the conjugate complex eigenvalues.
\end{itemize}
Let us denote such vectors with a superposed hat, and let us call them for simplicity (left and right) autovectors.
\end{definition}

\begin{lemma}\label{lemma:partialkgeneral}
Let $T$ be an $n\times n$ lower triangular block real matrix of the form (\ref{matrix_general}) 
where
$T^i_j$ are $n_i\times n_j$ matrices, and 
$0^i_j$  $n_i\times n_j$ matrices of zeros. 

The entries of matrices $T^i_j$ $(i=1,\ldots,k,\,j=1,\ldots,i)$ 
depend at most on the $m_i$ variables of the set $\mathcal{U}_i$ if and only if:
\begin{enumerate} 
\item the set of the eigenvalues of $T$ (counted with their multiplicity) with corresponding left and right autovectors can be divided
into $k$ subsets each containing $n_i$ $(i=1,\ldots,k)$ elements 
\begin{equation}\label{LambdaLR_general}
\begin{aligned}
&\left\{\{\Lambda^{(1,1)},\ldots,\Lambda^{(1,n_1)}\},\ldots,\{\Lambda^{(k,1)},\ldots,\Lambda^{(k,n_k)}\}\right\},\\
&\left\{\{\widehat{\mathbf{L}}^{(1,1)},\ldots,\widehat{\mathbf{L}}^{(1,n_1)}\},\ldots,\{\widehat{\mathbf{L}}^{(k,1)},\ldots,\widehat{\mathbf{L}}^{(k,n_k)}\}\right\},\\
&\left\{\{\widehat{\mathbf{R}}^{(1,1)},\ldots,\widehat{\mathbf{R}}^{(1,n_1)}\},\ldots,\{\widehat{\mathbf{R}}^{(k,1)},\ldots,\widehat{\mathbf{R}}^{(k,n_k)}\}\right\};
\end{aligned}
\end{equation}
\item the following structure conditions hold true:
\begin{equation}
\label{condgeneralized}
\begin{aligned} 
&\left(\nabla_{\mathbf{U}}\Lambda^{(i,\alpha)}\right)\cdot \widehat{\mathbf{R}}^{(j,\gamma)}=0,\\
&\widehat{\mathbf{L}}^{(i,\alpha)}\cdot\left((\nabla_{\mathbf{U}}\widehat{\mathbf{R}}^{(\ell,\beta)}) \widehat{\mathbf{R}}^{(j,\gamma)}
\right)=0, \\
&i=1,\ldots, k-1, \; \ell=1,\ldots,i,\\ 
&\alpha=1,\ldots,n_i,\;\beta=1,\ldots,n_\ell, \;\alpha\neq\beta\;\hbox{if}\;i=\ell,\\
&j=i+1,\ldots,k, \; \gamma=1,\ldots,n_j, 
\end{aligned}
\end{equation}
where 
\[
\nabla_{\mathbf{U}}\equiv\left(\frac{\partial}{\partial U^{(1,1)}},\ldots,\frac{\partial}{\partial U^{(1,n_1)}},\ldots,\frac{\partial}{\partial U^{(k,1)}},\ldots,\frac{\partial}{\partial U^{(k,n_k)}}\right).
\]
\end{enumerate}
\end{lemma}
\begin{proof}
The proof is as that of Lemma~\ref{lemma:partialk} taking into account that:
\begin{enumerate}
\item the left autovectors $\widehat{\mathbf{L}}^{(r,\alpha)}$ $(\alpha=1,\ldots,n_r)$ may have non--vanishing only the first $m_r$ components;
\item the right autovectors $\widehat{\mathbf{R}}^{(r,\alpha)}$ $(\alpha=1,\ldots,n_r)$
for $r>1$ may have non--vanishing only the last $n-m_{r-1}$ components.
\end{enumerate}

Moreover, using (\ref{condgeneralized}) and the
relations defining the generalized eigenvectors, it is also proved that the entries of matrices $T^i_r$ $(r=1,\ldots,i)$ are independent of the elements in the set $\overline{\mathcal{U}}_i$. 
\end{proof} 

Because of Lemma~\ref{lemma:partialkgeneral}, Theorem~\ref{thm:partialk} can be generalized to all first order autonomous and homogeneous quasilinear systems. 
\begin{theorem}\label{thm:partialkgeneral}
The first order quasilinear system (\ref{system1}) (not necessarily hyperbolic)
can be transformed by a smooth (locally) invertible transformation 
\begin{equation}
\label{transf_kgeneral}
\mathbf{u}=\mathbf{h}(\mathbf{U}),\quad \hbox{or, equivalently,}\quad
\mathbf{U}=\mathbf{H}(\mathbf{u}),
\end{equation}
into a system like
\begin{equation}\label{target_general}
\frac{\partial\mathbf{U}}{\partial t}+T(\mathbf{U})\frac{\partial\mathbf{U}}{\partial x}=\mathbf{0},
\end{equation}
in the unknowns (\ref{groupU}), where 
$T=(\nabla_{\mathbf{u}}\mathbf{H})\,A\,(\nabla_{\mathbf{u}}\mathbf{H})^{-1}$ is  a lower triangular block matrix having the form (\ref{matrix_general}),
with $T^i_j$ $(i=1,\ldots,k,\,j=1,\ldots,i)$  $n_i\times n_j$ matrices such that their entries are smooth functions depending at most on the elements of the set $\mathcal{U}_i$, 
whereas $0_j^i$ are $n_i\times n_j$ matrices of zeros, respectively, if and only if:
\begin{enumerate}
\item the set of the eigenvalues (counted with their multiplicity) of matrix $A$,  and the associated left and right autovectors can be divided into $k$ subsets each containing $n_i$ $(i=1,\ldots,k)$ elements, say  
\begin{equation}\label{lambdalr_general}
\begin{aligned}
&\left\{\{\lambda^{(1,1)},\ldots,\lambda^{(1,n_1)}\},\ldots,\{\lambda^{(k,1)},\ldots,\lambda^{(k,n_k)}\}\right\},\\
&\left\{\{\widehat{\mathbf{l}}^{(1,1)},\ldots,\widehat{\mathbf{l}}^{(1,n_1)}\},\ldots,\{\widehat{\mathbf{l}}^{(k,1)},\ldots,\widehat{\mathbf{l}}^{(k,n_k)}\}\right\},\\
&\left\{\{\widehat{\mathbf{r}}^{(1,1)},\ldots,\widehat{\mathbf{r}}^{(1,n_1)}\},\ldots,\{\widehat{\mathbf{r}}^{(k,1)},\ldots,\widehat{\mathbf{r}}^{(k,n_k)}\}\right\};
\end{aligned}
\end{equation} 
\item the following structure conditions hold true:
\begin{equation}\label{structureconditions_general}
\begin{aligned}
&\left(\nabla_{\mathbf{u}}\lambda^{(i,\alpha)}\right)\cdot \widehat{\mathbf{r}}^{(j,\gamma)}=0, \\
&\widehat{\mathbf{l}}^{(i,\alpha)}\cdot\left((\nabla_{\mathbf{u}}\widehat{\mathbf{r}}^{(\ell,\beta)}) \widehat{\mathbf{r}}^{(j,\gamma)}-
(\nabla_{\mathbf{u}}\widehat{\mathbf{r}}^{(j,\gamma)}) \widehat{\mathbf{r}}^{(\ell,\beta)}\right)=0, \\
&i=1,\ldots, k-1, \; \ell=1,\ldots,i,\\ 
&\alpha=1,\ldots,n_i,\;\beta=1,\ldots,n_\ell, \;\alpha\neq\beta\;\hbox{if}\;i=\ell,\\
&j=i+1,\ldots,k, \; \gamma=1,\ldots,n_j.
\end{aligned}
\end{equation}
\end{enumerate}
Moreover, the decoupling variables $U^{(i,\alpha)}=H^{(i,\alpha)}(\mathbf{u})$ 
$(i=1,\ldots,k-1, \,\alpha=1,\ldots,n_i)$ are found from 
\begin{equation}\label{transfconditions_general}
\left(\nabla_\mathbf{u}H^{(i,\alpha)}\right)\cdot \widehat{\mathbf{r}}^{(j,\gamma)}=0,
\end{equation}
where $j=i+1,\ldots,k, \,\gamma=1,\ldots,n_j$.
\end{theorem}

\begin{proof}
The proof, due to Lemma~\ref{lemma:partialkgeneral}, is as that of Theorem~\ref{thm:partialk}. 
\end{proof}

Finally, we are able to state the following theorem providing the conditions for the full decoupling of general first order quasilinear systems.

\begin{theorem}[Full decoupling of general systems]\label{thm:fullgeneral}
For the homogeneous and autonomous quasilinear system (\ref{system1}) (not necessarily hyperbolic) 
to be locally reducible into $k$
non--interacting subsystems of some orders $n_1,\ldots, n_k$, with $n_1 +\ldots+n_k=n$, in the unknowns (\ref{groupU}),
it is necessary and sufficient that: 
\begin{enumerate}
\item the eigenvalues of matrix $A$ (counted with their multiplicity), and the corresponding left and right autovectors can be divided into $k$ subsets  each containing $n_i$ $(i=1,\ldots,k)$ elements  as in (\ref{lambdalr_general}); 
\item the following  structure conditions hold true:
\begin{equation}
\begin{aligned}
&\left(\nabla_{\mathbf{u}}\lambda^{(i,\alpha)}\right)\cdot \widehat{\mathbf{r}}^{(j,\gamma)}=0, \\
&\widehat{\mathbf{l}}^{(i,\alpha)}\cdot\left((\nabla_{\mathbf{u}}\widehat{\mathbf{r}}^{(i,\beta)}) \widehat{\mathbf{r}}^{(j,\gamma)}-
(\nabla_{\mathbf{u}}\widehat{\mathbf{r}}^{(j,\gamma)}) \widehat{\mathbf{r}}^{(i,\beta)}\right)=0, \\
&\forall\,i,j=1,\ldots, k,\quad i\neq j,\\
&\alpha,\beta=1,\ldots,n_i, \quad \alpha\neq\beta, \quad \gamma=1,\ldots,n_j.
\end{aligned}
\end{equation}
\end{enumerate}
Moreover, the decoupling variables 
\[
U^{(i,\alpha)}=H^{(i,\alpha)}(\mathbf{u}), 
\]
are found from 
\begin{equation}
\left(\nabla_\mathbf{u}H^{(i,\alpha)}\right)\cdot \widehat{\mathbf{r}}^{(j,\gamma)}=0,
\end{equation}
where
\[
i,j=1,\ldots,k, \quad i\neq j,\quad \alpha=1,\ldots,n_i,\quad \gamma=1,\ldots,n_j.
\]
The coefficient matrix for a fully decoupled system results in block diagonal form (diagonal if $k=n$).
\end{theorem}

\begin{proof}
It immediately follows from Theorem~\ref{thm:partialkgeneral}.
\end{proof}

\section{Decoupling of nonhomogeneous and/or nonautonomous quasilinear systems}\label{main3}
In some physical applications it may occur to consider systems 
involving source terms, and/or systems where the coefficients may depend also on the independent variables, accounting for material inhomogeneities, or special geometric assumptions, or external 
actions \cite{Cristescu,Jeffrey1982,RogersRuggeri1985,Anliker1971,Pedley1980,CurroOliveri2008,Oliveri2012}. Therefore, one has to deal with nonhomogeneous and/or nonautonomous first order quasilinear systems of the form (\ref{system2}).

In some cases, systems like (\ref{system2}) may be transformed by a (locally) invertible point transformation to autonomous and homogeneous form or only to autonomous form preserving the 
quasilinear structure. This is possible if and only if the system (\ref{system2}) admits suitable algebras of Lie point symmetries.  In \cite{Oliveri2012} it has been proved a theorem stating necessary and sufficient conditions in order to map systems like (\ref{system2}) to autonomous and homogeneous form.
By relaxing the hypotheses, the same theorem can be used to map systems (\ref{system2}) into autonomous and nonhomogeneous first order quasilinear systems \cite{DonatoOliveri:1996,Oliveri:2010}. 

Therefore, three different situations may occur:
\begin{enumerate} 
\item System (\ref{system2}) can be mapped by an invertible point transformation to an equivalent autonomous and homogeneous first order quasilinear system in the independent variables 
$\widehat{t}(t,x)$, $\widehat{x}(t,x)$ and the dependent variables $\mathbf{U}=\mathbf{U}(t,x,\mathbf{u})$. It is required that it admits as subalgebra of its algebra of Lie point symmetries a three--dimensional Lie algebra spanned by the vector fields
\begin{equation}
\Xi_i=\tau_i(t,x)\frac{\partial}{\partial t}+\xi_i(t,x)\frac{\partial}{\partial x}+\sum_{A=1}^n\eta_i^A(t,x,\mathbf{u})\frac{\partial}{\partial u_A},\;\;(i=1,\ldots, 3),
\end{equation}
such that
\begin{equation}
\left[\Xi_i,\Xi_j\right]=0, \qquad \left[\Xi_i,\Xi_{3}\right]=\Xi_i, \qquad i,j=1,2;
\end{equation}
moreover, it has to be verified that all minors of order two extracted from the $3\times 2$ 
matrix with rows $(\tau_i,\xi_i)$ $(i=1,\ldots,3)$ are non--vanishing,
and the variables $\mathbf{U}$, which by construction are invariants of $\Xi_1$ and $\Xi_2$, result invariant with respect to $\Xi_3$ too. 

\item System (\ref{system2}) can be transformed to an equivalent autonomous and nonhomogeneous first order quasilinear system in the independent variables 
$\widehat{t}(t,x)$, $\widehat{x}(t,x)$ and the dependent variables $\mathbf{U}=\mathbf{U}(t,x,\mathbf{u})$. It is required that it  admits as subalgebra of its algebra of Lie point symmetries a two--dimensional Lie algebra spanned by the vector fields
\begin{equation}
\Xi_i=\tau_i(t,x)\frac{\partial}{\partial t}+\xi_i(t,x)\frac{\partial}{\partial x}+\sum_{A=1}^n\eta_i^A(t,x,\mathbf{u})\frac{\partial}{\partial u_A},\;\;(i=1, 2),
\end{equation}
such that the $2\times 2$ matrix with rows $(\tau_i,\xi_i)$ $(i=1,2)$ is non--singular and
\begin{equation}
\left[\Xi_i,\Xi_j\right]=0, \qquad i,j=1,2. 
\end{equation}

\item System (\ref{system2}) can not be transformed to autonomous form.

\end{enumerate}

In the first case, the decoupling problem can be faced by using the results of previous Sections, whereas cases 2 and 3 can be managed together.

It is worth of being observed that the decoupling of the system (\ref{system2}) is not affected
by a variable change of the independent variables provided that the new independent variables depend only on the old independent variables. Therefore, we can manage in a unified way nonhomogeneous quasilinear systems either when they are autonomous or not. So, we introduce only new dependent variables $\mathbf{U}$, as suitable functions of the old dependent variables, and state the following two theorems for the partial and the full decoupling. 

\begin{theorem}[Partial decoupling for quasilinear systems]\label{thm:partialkgeneral2}
The first order quasilinear system (\ref{system2})
can be transformed by a smooth (locally) invertible transformation 
\begin{equation}
\label{transf_kgeneral2}
\mathbf{u}=\mathbf{h}(\mathbf{U}),\quad \hbox{or, equivalently,}\quad
\mathbf{U}=\mathbf{H}(\mathbf{u}),
\end{equation}
into a system like
\begin{equation}\label{target_general2}
\frac{\partial\mathbf{U}}{\partial t}+T(t,x,\mathbf{U})\frac{\partial\mathbf{U}}{\partial x}=\mathbf{G}(t,x,\mathbf{U}),
\end{equation}
in the unknowns (\ref{groupU}), where 
\begin{equation}
\mathbf{G}\equiv\left(G^{(1,1)},\ldots,G^{(1,n_1)},\ldots,G^{(k,1)},\ldots,G^{(k,n_k)}\right)^T,
\end{equation}
$T=(\nabla_{\mathbf{u}}\mathbf{H})\,A\,(\nabla_{\mathbf{u}}\mathbf{H})^{-1}$ being  a lower triangular block matrix having the form (\ref{matrix_general}), $\mathbf{G}=(\nabla_{\mathbf{u}}\mathbf{H})\mathbf{g}$,
such that $T^i_j$ and $G^{(i,\alpha)}$ $(i=1,\ldots,k,\,j=1,\ldots,i,\,\alpha=1,\ldots,n_i)$ depend at most on $t$, $x$ and the elements of the set  $\mathcal{U}_i$, 
whereas $0_j^i$ are $n_i\times n_j$ matrices of zeros, respectively,
if and only if:
\begin{enumerate} 
\item the set of the eigenvalues of matrix $A$ (counted with their multiplicity), and the associated left and right autovectors can be divided
into $k$ subsets as in (\ref{lambdalr_general});
\item the following structure conditions hold true:
\begin{equation}\label{structure_conditionsgeneral2}
\begin{aligned}
&\left(\nabla_{\mathbf{u}}\lambda^{(i,\alpha)}\right)\cdot \widehat{\mathbf{r}}^{(j,\gamma)}=0, \\
&\widehat{\mathbf{l}}^{(i,\alpha)}\cdot\left((\nabla_{\mathbf{u}}\widehat{\mathbf{r}}^{(\ell,\beta)}) \widehat{\mathbf{r}}^{(j,\gamma)}-
(\nabla_{\mathbf{u}}\widehat{\mathbf{r}}^{(j,\gamma)}) \widehat{\mathbf{r}}^{(\ell,\beta)}\right)=0, \\
&\left(\nabla_\mathbf{u}(\widehat{\mathbf{l}}^{(i,\alpha)}\cdot \mathbf{g})\right)\cdot\widehat{\mathbf{r}}^{(j,\gamma)}=0,\\
&i=1,\ldots, k-1, \; \ell=1,\ldots,i,\\ 
&\alpha=1,\ldots,n_i,\;\beta=1,\ldots,n_\ell, \;\alpha\neq\beta\;\hbox{if}\;i=\ell,\\
&j=i+1,\ldots,k, \; \gamma=1,\ldots,n_j.
\end{aligned}
\end{equation}
\end{enumerate}
Moreover, the decoupling variables $U^{(i,\alpha)}=H^{(i,\alpha)}(\mathbf{u})$ 
$(i=1,\ldots,k-1, \,\alpha=1,\ldots,n_i)$ are found from 
\begin{equation}\label{transfconditions_general2}
\left(\nabla_\mathbf{u}H^{(i,\alpha)}\right)\cdot \widehat{\mathbf{r}}^{(j,\gamma)}=0,
\end{equation}
where $j=i+1,\ldots,k, \,\gamma=1,\ldots,n_j$.
\end{theorem}

\begin{proof}
The proof, due to Lemma~\ref{lemma:partialkgeneral}, follows the same steps  as those of Theorem~\ref{thm:partialk}, the only difference being in the additional requirement expressed by (\ref{structure_conditionsgeneral2})$_3$. 

It is
\begin{equation}
\begin{aligned}
0&=\left(\nabla_\mathbf{u}(\widehat{\mathbf{l}}^{(i,\alpha)}\cdot\mathbf{g})\right)\cdot \widehat{\mathbf{r}}^{(j,\gamma)}=\\
&=\left(\nabla_\mathbf{U}(\widehat{\mathbf{l}}^{(i,\alpha)}\cdot\mathbf{g})\right) (\nabla_\mathbf{u}\mathbf{H})(\nabla_\mathbf{u}\mathbf{H})^{-1}\widehat{\mathbf{R}}^{(j,\gamma)}=\\
&=\left(\nabla_\mathbf{U}(\widehat{\mathbf{L}}^{(i,\alpha)}(\nabla_\mathbf{u}\mathbf{H})(\nabla_\mathbf{u}\mathbf{H})^{-1}\mathbf{G})\right)\cdot \widehat{\mathbf{R}}^{(j,\gamma)}=\\
&=\left(\nabla_\mathbf{U}(\widehat{\mathbf{L}}^{(i,\alpha)}\cdot\mathbf{G})\right)\cdot \widehat{\mathbf{R}}^{(j,\gamma)}. 
\end{aligned}
\end{equation}
Therefore,
\begin{equation}
\left(\nabla_\mathbf{u}(\widehat{\mathbf{l}}^{(i,\alpha)}\cdot\mathbf{g})\right)\cdot \widehat{\mathbf{r}}^{(j,\gamma)}=0\quad\Leftrightarrow\quad
\left(\nabla_\mathbf{U}(\widehat{\mathbf{L}}^{(i,\alpha)}\cdot\mathbf{G})\right)\cdot \widehat{\mathbf{R}}^{(j,\gamma)}=0.
\end{equation}

Since $\widehat{\mathbf{L}}^{(i,\alpha)}$ may have non--vanishing only the first $m_i$ components, whereas $\widehat{\mathbf{R}}^{(j,\gamma)}$ may have non--vanishing only the last $n-m_{j-1}$ components, conditions
\begin{equation}
\left(\nabla_\mathbf{U}(\widehat{\mathbf{L}}^{(i,\alpha)}\cdot\mathbf{G})\right)\cdot \widehat{\mathbf{R}}^{(j,\gamma)}=0
\end{equation}
are necessary and sufficient in order the components $\{\mathbf{G}^{(r,1)},\ldots,\mathbf{G}^{(r,n_r)}\}$ to be dependent at most on $t$, $x$ and
the elements of the set ${\mathcal{U}}_r$, and this concludes the proof.
\end{proof}

Finally, we are able to state the following theorem providing a solution to the full decoupling problem for general nonhomogeneous and/or nonautonomous first order quasilinear systems.

\begin{theorem}[Full decoupling for nonhomogeneous quasilinear systems]
For the  first order nonhomogeneous and/or nonautonomous quasilinear
system (\ref{system2})
to be locally reducible into $k$ non--interacting subsystems of some orders $n_1,\ldots, n_k$, with $n_1+\ldots+n_k=n$, in the unknowns (\ref{groupU}),
it is necessary and sufficient that: 
\begin{enumerate}
\item the eigenvalues of matrix $A$ (counted with their multiplicity), and the corresponding left and right autovectors can be divided
into $k$ subsets each containing $n_i$ $(i=1,\ldots,k)$ elements as in (\ref{lambdalr_general}); 
\item the following structure conditions hold true:
\begin{equation}
\begin{aligned}
&\left(\nabla_{\mathbf{u}}\lambda^{(i,\alpha)}\right)\cdot \widehat{\mathbf{r}}^{(j,\gamma)}=0, \\
&\widehat{\mathbf{l}}^{(i,\alpha)}\cdot\left((\nabla_{\mathbf{u}}\widehat{\mathbf{r}}^{(i,\beta)}) \widehat{\mathbf{r}}^{(j,\gamma)}-
(\nabla_{\mathbf{u}}\widehat{\mathbf{r}}^{(j,\gamma)}) \widehat{\mathbf{r}}^{(i,\beta)}\right)=0,\\
&\left(\nabla_\mathbf{u}(\widehat{\mathbf{l}}^{(i,\alpha)}\cdot \mathbf{g})\right)\cdot\widehat{\mathbf{r}}^{(j,\gamma)}=0,\\
&\forall\,i,j=1,\ldots, k,\quad i\neq j,\\
&\alpha,\beta=1,\ldots,n_i, \quad \alpha\neq\beta, \quad \gamma=1,\ldots,n_j.
\end{aligned}
\end{equation}
\end{enumerate}
Moreover, the decoupling variables 
\[
U^{(i,\alpha)}=H^{(i,\alpha)}(\mathbf{u}), 
\]
are found from 
\begin{equation}
\left(\nabla_\mathbf{u}H^{(i,\alpha)}\right)\cdot \widehat{\mathbf{r}}^{(j,\gamma)}=0,
\end{equation}
where
\[
i,j=1,\ldots,k, \quad i\neq j,\quad \alpha=1,\ldots,n_i,\quad \gamma=1,\ldots,n_j.
\]
\end{theorem}

\begin{proof}
It immediately follows from Theorem~\ref{thm:partialkgeneral2}.
\end{proof}

\section{Applications}\label{applications}
In this Section, we consider some applications of the results above derived. As far as the notation is concerned,  the components of the field $\mathbf{u}$ are denoted with the symbols
typically used in the applications. 

The first two examples are related to the Euler equation of an ideal gas with the special value $\Gamma=3$ \cite[p.~88]{Courant-Friedrichs} for the adiabatic index, whereas the third example concerns the equations of a model of travelling threadline with a particular constitutive law for the tension.

\begin{example}{One--dimensional Euler equations of barotropic fluids.}

Let us consider the one--dimensional Euler equations of a barotropic fluid
\begin{equation}
\frac{\partial\mathbf{u}}{\partial t}+A(\mathbf{u})\frac{\partial\mathbf{u}}{\partial x}=\mathbf{0},
\end{equation}
with
\begin{equation}\label{Eul}
\mathbf{u}=\left[
\begin{array}{c}
\rho \\v
\end{array}
\right], \qquad
A=\left[
\begin{array}{cc}
v & \rho\\
\frac{p^\prime(\rho)}{\rho} & v
\end{array}
\right],
\end{equation}
where $\rho(t,x)$ is the mass density, $v(t,x)$ the velocity, and $p(\rho)$ the pressure. This system is strictly hyperbolic provided that $p^\prime(\rho)>0$ (the prime denoting the differentiation with respect to the argument), with characteristic velocities
\begin{equation}
\lambda_{1,2}= v\pm\sqrt{p^{\prime}(\rho)},
\end{equation}
to which correspond the left and right eigenvectors
\begin{equation}
\begin{aligned}
&\mathbf{l}_{1,2}=\left(\sqrt{p^\prime(\rho)},\pm\rho\right), \qquad 
\mathbf{r}_{1,2}=\left(
\begin{array}{c}
\rho \\ \pm\sqrt{p^\prime(\rho)}
\end{array}
\right).
\end{aligned}
\end{equation}
The conditions for the possible decoupling provide the constraint
\begin{equation}
\rho p^{\prime\prime}(\rho)-2p^\prime(\rho)=0,
\end{equation}
which is satisfied by the special constitutive law
\begin{equation}
p(\rho)=p_0\rho^3, \quad p_0\;\hbox{constant}.
\end{equation}
Thus, if the adiabatic index is equal to 3 (in this case the characteristics are straight lines, \cite[p.~88]{Courant-Friedrichs}),  we may introduce the variable transformation
\begin{equation}
\mathbf{U}=\mathbf{H}(\mathbf{u})
\end{equation}
such that
\begin{equation}
\left(\nabla_{\mathbf{u}}H_1\right)\cdot \mathbf{r}_2=0,\qquad \left(\nabla_{\mathbf{u}}H_2\right)\cdot \mathbf{r}_1=0.
\end{equation}
As a consequence, by choosing
\begin{equation}
U_1=H_1(\rho,v)=v+\sqrt{3p_0}\rho,\,\qquad U_2=H_2(\rho,v)=v-\sqrt{3p_0} \rho,
\end{equation}
we obtain the following fully decoupled system
\begin{equation}
\begin{aligned}
&\frac{\partial U_1}{\partial t}+U_1\frac{\partial U_1}{\partial x}=0,\\
&\frac{\partial U_2}{\partial t}+U_2\frac{\partial U_2}{\partial x}=0.
\end{aligned}
\end{equation}
\end{example}

\begin{example}{One--dimensional isentropic gas dynamics equations.}

Let us consider the one--dimensional Euler equations for the isentropic flow of an ideal fluid subject to no external forces,
\begin{equation}
\frac{\partial\mathbf{u}}{\partial t}+A(\mathbf{u})\frac{\partial\mathbf{u}}{\partial x}=\mathbf{0},
\end{equation}
with
\begin{equation}
\mathbf{u}=\left[
\begin{array}{c}
\rho \\ v\\ s\\
\end{array}
\right], \qquad
A=\left[
\begin{array}{ccc}
v & \rho & 0\\
\frac{1}{\rho}\frac{\partial p}{\partial \rho} & v & \frac{1}{\rho}\frac{\partial p}{\partial s}\\
0 & 0 & v\\
\end{array}
\right],
\end{equation}
where $\rho(t,x)$ is the mass density, $v(t,x)$ the velocity, $s(t,x)$ the entropy, and $p(\rho,s)$ the pressure. 

The eigenvalues of matrix $A$ are
\begin{equation}
\lambda_{1,2}= v\pm\sqrt{\frac{\partial p}{\partial \rho}},\qquad \lambda_3= v,
\end{equation}
with associated left and right eigenvectors
\begin{equation}
\begin{aligned}
\mathbf{l}_{1,2}&=\left(\sqrt{\frac{\partial p}{\partial \rho}},\pm\rho, \frac{\rho}{s} \sqrt{\frac{\partial p}{\partial \rho}}\right), \qquad &&\mathbf{l}_3= (0, 0, 1),\\
\mathbf{r}_{1,2}&=\left(
\begin{array}{c}
\rho \\ \pm\sqrt{\frac{\partial p}{\partial \rho}}\\ 0
\end{array}
\right), \qquad &&\mathbf{r}_{3}=
\left(
\begin{array}{c}
\rho \\ 0\\ -s\\
\end{array}
\right).
\end{aligned}
\end{equation}
The constraints
\begin{equation}
\begin{aligned}
&\left(\nabla_\mathbf{u}\lambda_1\right)\cdot \mathbf{r}_{2}=0,\qquad \left(\nabla_\mathbf{u}\lambda_1\right)\cdot \mathbf{r}_{3}=0,\\
&\left(\nabla_\mathbf{u}\lambda_2\right)\cdot \mathbf{r}_{1}=0,\qquad \left(\nabla_\mathbf{u}\lambda_2\right)\cdot \mathbf{r}_{3}=0,
\end{aligned}
\end{equation}
are satisfied with the constitutive law
\begin{equation}
p(\rho,s)=p_0\rho^3 s^2+f(s), 
\end{equation}
where $p_0$ is constant and $f(s)$ a function of its argument; therefore, we may introduce the variable transformation
\begin{equation}
\mathbf{U}=\mathbf{H}(\mathbf{u})
\end{equation}
such that
\begin{equation}
\begin{aligned}
&\left(\nabla_{\mathbf{u}}H_1\right)\cdot \mathbf{r}_{2}=0,\qquad \left(\nabla_{\mathbf{u}}H_1\right)\cdot \mathbf{r}_{3}=0,\\
&\left(\nabla_{\mathbf{u}}H_2\right)\cdot \mathbf{r}_{1}=0,\qquad \left(\nabla_{\mathbf{u}}H_2\right)\cdot \mathbf{r}_{3}=0.
\end{aligned}
\end{equation}
As a consequence, by choosing
\begin{equation}
\begin{aligned}
&U_1=H_1(\rho,v,s)=v+\sqrt{3p_0}\rho s,\\
&U_2=H_2(\rho,v,s)=v-\sqrt{3p_0}\rho s,\\
&U_3=H_3(\rho,v,s)=s,
\end{aligned}
\end{equation}
we obtain the following partially decoupled system
\begin{equation}
\begin{aligned}
&\frac{\partial U_1}{\partial t}+U_1\frac{\partial U_1}{\partial x}=0,\\ 
&\frac{\partial U_2}{\partial t}+U_2\frac{\partial U_2}{\partial x}=0,\\
&\frac{\partial U_3}{\partial t}+\frac{1}{2}(U_1+U_2)\frac{\partial U_3}{\partial x}=0,
\end{aligned}
\end{equation}
where the first two equations can be solved independently from each other and the third one.
\end{example}

\begin{example}{Model of travelling threadline.}

Let us consider the nonlinear model describing the motion of a moving threadline \cite{AmesLeeZaiser:1968,DonatoOliveri:1988}
taking into account both geometric and material nonlinearities.

Based upon the following hypotheses:
\begin{itemize}
\item the motion is two--dimensional;
\item the string is elastic and always in tension;
\item the string is perfectly flexible;
\item the effects of gravity and air drag are neglected,
\end{itemize}
in \cite{AmesLeeZaiser:1968} the following equations have been derived:
\begin{equation}
\begin{aligned}
&m(1+u_x^2)^{1/2}\frac{dV^x}{dt}=\frac{\partial}{\partial x}\left(T\sin\theta\right),\\
&m(1+u_x^2)^{1/2}\frac{dV^y}{dt}=\frac{\partial}{\partial x}\left(T\cos\theta\right),
\end{aligned}
\end{equation}
supplemented by the continuity equation
\begin{equation}
\frac{d}{dt}\left(m(1+u_x^2)^{1/2}\right)+m(1+u_x^2)^{1/2} \frac{\partial V^x}{\partial x}=0,
\end{equation}
and a constitutive law in the form
\begin{equation}
T=T(m,m_t).
\end{equation}
In the previous equations, $m$ is the mass per unit length, $V^x$ the axial
component of the velocity, $V^y$ the transverse component of the velocity, $T$  the
tension, and $u$ the transverse displacement; moreover, the subscripts $t$ and $x$ denote partial derivatives with
respect to the indicated variables.

Upon introduction of the quantity
\begin{equation}
\rho = m (1 + u_x^2)^{1/2},
\end{equation}
and taking into account (see \cite{AmesLeeZaiser:1968} for details) that
\begin{equation}
\begin{aligned}
&\sin\theta = \frac{u_x}{(1+u_x^2)^{1/2}}, \qquad \cos\theta = \frac{1}{(1+u_x^2)^{1/2}}, \\
&V^y=u_t+V^xu_x=v+V^x\varepsilon,
\end{aligned}
\end{equation}
assuming the constitutive equation for the tension in the form $T=T(m)$ \cite{DonatoOliveri:1988}, the governing equation can be written as
\begin{equation}
\label{threadline}
\frac{\partial \mathbf{u}}{\partial t}+A(\mathbf{u})\frac{\partial\mathbf{u}}{\partial x}=\mathbf{0},
\end{equation}
where
\begin{equation}
\mathbf{u}=\left(
\begin{array}{c}
\rho \\ V^x \\ v  \\ \varepsilon
\end{array}
\right), \qquad 
A=\left[
\begin{array}{cccc}
V^x & \rho & 0 & 0 \\
\frac{-T^\prime}{\rho(1+\varepsilon^2)} & V^x & 0 & \frac{\varepsilon}{1+\varepsilon^2)^2}\left(T^\prime+\frac{T}{m}\right)\\
0 & 0 & 2V^x & (V^x)^2-\frac{T}{m(1+\varepsilon^2)}\\
0 & 0 & -1 & 0
\end{array}
\right].
\end{equation}
The eigenvalues of matrix $A$ are
\begin{equation}
\lambda_{1,2}=V^x\pm \left(\frac{-T^\prime}{1+\varepsilon^2)}\right)^{1/2},\qquad \lambda_{3,4}=V^x\pm\left(\frac{T}{m(1+\varepsilon^2)}\right)^{1/2},
\end{equation}
with associated left and right eigenvectors
\begin{equation}
\begin{aligned}
&\mathbf{l}_{1,2}=\left(\pm \frac{\sqrt{-(1+\epsilon^2) T^{\prime}}}{\rho\epsilon\left(V^x\mp\sqrt{\frac{-T^\prime}{1+\epsilon^2}}\right)},\frac{1+\epsilon^2}{\epsilon\left(V^x\mp\sqrt{\frac{-T^\prime}{1+\epsilon^2}}\right)},\frac{1}{V^x\mp\sqrt{\frac{-T^\prime}{1+\epsilon^2}}},1\right),\\
&\mathbf{l}_{3,4}=\left(0, 0,\rho, \rho V^x\pm\sqrt{\frac{\rho  T}{(1+\epsilon^2)^{1/2}}}\right),\\
&\mathbf{r}_{1,2}=\left(
\begin{array}{c}
\rho \\ \pm\left(\frac{-T^\prime}{1+\varepsilon^2}\right)^{1/2} \\ 0 \\0
\end{array}
\right), \qquad
\mathbf{r}_{3,4}=\left(
\begin{array}{c}
\frac{\rho\varepsilon}{1+\varepsilon^2} \\ 
\pm\left(\frac{T}{m(1+\varepsilon^2)}\right)^{1/2}\frac{\varepsilon}{1+\varepsilon^2} \\
-\left(V^x\pm\frac{T}{m(1+\varepsilon^2)}\right)^{1/2} \\1
\end{array}
\right). 
\end{aligned}
\end{equation}
The structure conditions for the partial decoupling
\begin{equation}
\begin{aligned}
&(\nabla_\mathbf{u}\lambda_i)\cdot \mathbf{r}_j=0,\\
&\mathbf{l}_{i}\cdot\left((\nabla_{\mathbf{u}}\mathbf{r}_{\ell}) \mathbf{r}_{j}-
(\nabla_{\mathbf{u}}\mathbf{r}_{j}) \mathbf{r}_{\ell}\right)=0, \qquad i,\ell=1,2,\quad i\neq\ell, \quad j=3,4, \\
\end{aligned}
\end{equation}
are satisfied with the following constitutive law
\begin{equation}
\label{constitutive}
T(m)=\frac{k}{m},\qquad k\;\hbox{constant}.
\end{equation}
Then, we may introduce the variable transformation
\begin{equation}
\mathbf{U}=\mathbf{H}(\mathbf{u})
\end{equation}
such that
\begin{equation}\label{condHr}
\left(\nabla_{\mathbf{u}}H_i\right)\cdot \mathbf{r}_{j}=0,\qquad  \qquad i,=1,2, \quad j=3,4.
\end{equation}
By integrating relations (\ref{condHr}), \textit{i.e},
\begin{equation}
\begin{aligned}
&\left(V^x+\frac{\sqrt{k}}{\rho}\right)\frac{\partial H_1}{\partial v}-\frac{\partial H_1}{\partial \epsilon}=0, && \left(V^x+\frac{\sqrt{k}}{\rho}\right)\frac{\partial H_2}{\partial v}-\frac{\partial H_2}{\partial \epsilon}=0,\\
&\left(V^x-\frac{\sqrt{k}}{\rho}\right)\frac{\partial H_1}{\partial v}-\frac{\partial H_1}{\partial \epsilon}=0, && \left(V^x-\frac{\sqrt{k}}{\rho}\right)\frac{\partial H_2}{\partial v}-\frac{\partial H_2}{\partial \epsilon}=0,\\
\end{aligned}
\end{equation}
it follows that
\begin{equation}
H_1=H_1\left(\rho,V^x\right),\qquad H_2=H_2\left(\rho,V^x\right).
\end{equation}
As a consequence, by choosing the identity transformation, we obtain this 
partially decoupled system
\begin{equation}
\left\{
\begin{aligned}
&\frac{\partial \rho}{\partial t}+V^x\frac{\partial \rho}{\partial x}+\rho\frac{\partial V^x}{\partial x}=0,\\
&\frac{\partial V^x}{\partial t}+\frac{k}{\rho^{3}}\frac{\partial \rho}{\partial x}+V^x\frac{\partial V^x}{\partial x}=0,\\
&\frac{\partial v}{\partial t}+2 V^x\frac{\partial v}{\partial x}+\left({(V^x)}^2-\frac{k}{\rho^2}\right)\frac{\partial \epsilon}{\partial x}=0,\\
&\frac{\partial \epsilon}{\partial t}-\frac{\partial v}{\partial x}=0.
\end{aligned}\right.
\end{equation}
It is worth of being observed that with the constitutive relation (\ref{constitutive}) the system (\ref{threadline}) has two distinct eigenvalues each with multiplicity 2, and is completely exceptional \cite{Lax,Boillat}.
\end{example}

\section*{Acknoledgments}
Work supported by ``Gruppo Nazionale per la Fisica Matematica'' of ``Istituto Nazionale di Alta Matematica''. The authors
are grateful to the unknown Referee whose comments allowed us to clarify some aspects so improving the quality of the paper.

\end{document}